\documentclass[conference]{IEEEtran}
\IEEEoverridecommandlockouts
\usepackage{cite}
\usepackage{amsmath,amssymb,amsfonts}
\usepackage{algorithmic}
\usepackage{graphicx}
\usepackage{textcomp}
\usepackage{xcolor}

\usepackage{subfigure}

\usepackage{flushend}

\def\BibTeX{{\rm B\kern-.05em{\sc i\kern-.025em b}\kern-.08em
    T\kern-.1667em\lower.7ex\hbox{E}\kern-.125emX}}
\begin{document}

\title{Probabilistic Constellation Shaping for OFDM-Based ISAC Signaling}

\author{\IEEEauthorblockN{Zhen Du$^{*\dagger}$, Fan Liu$^\dagger$, Yifeng Xiong$^\ddagger$, Tony Xiao Han$^\star$, Weijie Yuan$^\dagger$, Yuanhao Cui$^\dagger$, Changhua Yao$^{*}$, \\ Yonina C. Eldar$^\circ$}
\IEEEauthorblockA{$^*$\textit{School of Electronic and Information Engineering, Nanjing University of Information Science \& Technology, Nanjing, China } \\
$^\dagger$\textit{Department of Electronic and Electrical Engineering, Southern University of Science and Technology, Shenzhen, China}\\
$^\ddagger$\textit{School of Information and Electronic Engineering,	Beijing University of Posts and Telecommunications, Beijing, China}\\
$^\star$\textit{Huawei Technologies Ltd., Shenzhen, China}\\
$^\circ$\textit{Faculty of Mathematics and Computer Science, Weizmann Institute of Science, Rehovot, Israel}\\
Email: {duzhen@nuist.edu.cn, liuf6@sustech.edu.cn, yifengxiong@bupt.edu.cn, tony.hanxiao@huawei.com, } \\ {\{yuanwj, cuiyh\}@sustech.edu.cn, ych2347@163.com, yonina.eldar@weizmann.ac.il}}
}

\maketitle

\begin{abstract}
Integrated Sensing and Communications (ISAC) has garnered significant attention as a promising technology for the upcoming sixth-generation wireless communication systems (6G). In pursuit of this goal, a common strategy is that a unified waveform, such as Orthogonal Frequency Division Multiplexing (OFDM), should serve dual-functional roles by enabling simultaneous sensing and communications (S\&C) operations.
However, the sensing performance of an OFDM communication signal is substantially affected by the randomness of the data symbols mapped from bit streams. Therefore, achieving a balance between preserving communication capability (i.e., the randomness) while improving sensing performance remains a challenging task. To cope with this issue, in this paper we analyze the ambiguity function of the OFDM communication signal modulated by random data. Subsequently, a probabilistic constellation shaping (PCS) method is proposed to devise the probability distributions of constellation points, which is able to strike a scalable S\&C tradeoff of the random transmitted signal. Finally, the superiority of the proposed PCS method over conventional uniformly distributed constellations is validated through numerical simulations.
\end{abstract}


\section{Introduction}

The International Telecommunication Union (ITU) has recently granted official recognition to ISAC as one of the six key usage scenarios of 6G \cite{ITU-R}. In this paradigm, the traditional approach of treating S\&C functionalities as separate objectives is discarded. Instead, they are synergistically designed to achieve mutual benefits, driven by technological advancements and commercial demands \cite{liu2022integrated,du2023towards}. To that end, developing a unified waveform that enables simultaneous information transmission and target sensing becomes crucial. While OFDM is widely employed as the default waveform in 4G and 5G cellular communications, its capability of sensing remains to be developed in 6G networks \cite{sturm2011waveform}. A fundamental issue arisen in OFDM based ISAC is how to characterize the sensing performance when signals are embedded with random communication data, which may jeopardize the target detection performance for sensing. More importantly, how to control the randomness of the OFDM signal, such that a scalable performance tradeoff between S\&C can be achieved, still remains an open problem.

It is widely recognized that conventional radar waveforms require adherence to the constant modulus constraint, such as the frequency-modulated continuous-wave (FMCW) signals that are widely adopted for autonomous vehicles \cite{ma2020joint}. This constraint ensures a flat spectrum and facilitates narrow mainlobes and low sidelobes in the output of matched filtering. Phase shift keying (such as BPSK, QPSK, 8-PSK)-based OFDM schemes typically fulfill this requirement. However, higher-order quadrature amplitude modulation (such as 16-QAM, 64-QAM) fail to meet this criterion due to randomly varying amplitudes. Consequently, using the OFDM communication signal modulated with QAM symbols directly for sensing leads to compromised sensing performance \cite{xiong2023fundamental}. 
Such an issue becomes increasingly critical in various ISAC applications in 6G networks. One typical example is vehicle tracking in NR-V2X networks, where the utilization of legacy OFDM communication signals for matched filtering becomes essential in terms of range and Doppler estimation \cite{du2023towards,du2022integrated}.

%In \cite{b2}, a element-wise division is exploited to eliminate random amplitude and phase states obtained through any PSK or QAM modulation scheme, thereby obtaining a matched filtering output with the DFT structure of OFDM. However, such an operation unavoidably involves the random amplitudes and phases in the noise, which is still harmful to the estimation and detection.

To measure the sensing performance of random OFDM communication signals, we adopt the ambiguity function (AF) as a basic tool, which is defined as the two-dimensional correlation between the transmitted signal and its duplicated version subject to time-delay and frequency-shift \cite{sen2009adaptive}. Specifically, for random OFDM communication signals, we aim to analyze the statistical characteristics of their AFs, encompassing the expected value and variance of the matched filtering output. While previous studies, such as \cite{tigrek2012ofdm}, have explored these properties for PSK-OFDM, the implications for QAM-OFDM or other modulation schemes remain unexplored.

On top of the AF analysis, a more important task is to discover a new degree of freedom (DoF) in waveform design, allowing us to incorporate a flexible tradeoff between communication-centric and sensing-centric designs. 
To be more specific, we hope to reserve the communication capability (i.e., the achievable rate) of a given QAM-format OFDM signal, while improving its sensing ability.
We propose to realize this goal through a specifically tailored constellation shaping approach \cite{bocherer2014probabilistic,bocherer2015bandwidth,cho2019probabilistic,steiner2020coding}. Constellation shaping techniques may be split into two categories: probabilistic constellation shaping (PCS) and geometric constellation shaping (GCS). In communication systems, constellation shaping has been leveraged to minimize the gap between the achievable rate and Shannon capacity \cite{steiner2020coding}. In this paper, we focus on PCS design to improve the sensing performance for QAM-OFDM signals, while striking a scalable S\&C tradeoff in ISAC systems.

This paper begins by establishing a model for the OFDM communication signal and conducting an analysis of its AF. Subsequently, we highlight the key distinction between PSK and QAM by examining the variance of the AF. To balance between S\&C objectives, we incorporate the PCS approach to design a QAM-based constellation. Finally, we evaluate both S\&C performance of the proposed PCS approach through numerical simulations.

\section{Signal Model, Ambiguity Function and Statistical Characteristics}\label{sec2}
In this section, we commence with the OFDM signal model, followed by the analysis of the statistical characteristics of its AF. This lays the foundation for the proposed PCS approach in Sec. \ref{sec3}.

\subsection{Ambiguity Function}
We consider a single-symbol OFDM signal with $L$ subcarriers expressed as
\begin{equation}\label{equ1}
	\begin{aligned}
		s(t) = \sum^{L-1}_{l=0}A_l e^{j\psi_l} \underbrace{e^{j2\pi l \Delta ft}}_{\phi_l(t)} \text{rect}\left( \frac{t}{T_p} \right),
	\end{aligned}
\end{equation}
where $A_l$ and $\psi_l$ denote the amplitude and phase of the $l$th random symbol in the constellation, respectively; $T_p$ is the symbol duration; $\text{rect}(t)=1$ for $0\leq t\leq 1$, and zero otherwise. 
It is noted that the randomness of OFDM communication signals lies in the discrete random variables $A_l$ and $\psi_l$ in the given constellation. 
Assume that there are $Q$ discrete constellation points, where the $q$th point is transmitted with the probability of $p_q$, satisfying $\sum^Q_{q=1}p_q=1$. As a special case, conventional PSK and QAM constellations are with uniformly distributed points, i.e., $p_q = \frac{1}{Q}, \ \forall q$.
Then the expectation $E_X\{x\}=\sum^Q_{q=1}x_qp_q$ refers to the summation of $Q$ values weighted by their discrete probabilities in the constellation, where $x$ represents a function of the random constellation points $A_l e^{\psi_l}$. 
For convenience, we omit the subscript of $E_X\{x\}$ in the following. With this definition, the normalized transmit power can be expressed as $E\{A^2_l\} = 1$.

The AF of $s(t)$ can be then expressed as \cite{sen2009adaptive,tigrek2012ofdm}
\begin{equation}\label{equ2}
	\begin{aligned}
		\mathcal{X}(\tau,\nu) = & \int^{\infty}_{-\infty} s(t)s^*(t-\tau) e^{-j2\pi\nu t}dt \\
		= & \sum^{L-1}_{l_1=0}\sum^{L-1}_{l_2=0} A_{l_1}A_{l_2}e^{j(\psi_{l_1}-\psi_{l_2})} \\ & \cdot \int^{\infty}_{-\infty} \phi_{l_1}(t)\phi^*_{l_2}(t-\tau) e^{-j2\pi\nu t} dt.
	\end{aligned}
\end{equation}
Note that the integral in (\ref{equ2}) may be further recast as
\begin{equation}\label{equ3}
	\begin{aligned}
		& \int^{\infty}_{-\infty} \phi_{l_1}(t)\phi^*_{l_2}(t-\tau)  e^{-j2\pi\nu t} dt 
		\\ & = e^{j2\pi l_2\Delta f \tau} \cdot \int^{T_\text{max}}_{T_\text{min}} e^{j2\pi \left((l_1-l_2)\Delta f -\nu\right) t} dt,
	\end{aligned}
\end{equation}
where $T_\text{min}=\max(0,\tau)$, and $T_\text{max}=\min(T_p,T_p+\tau)$.

For notational simplifications, we denote
\begin{equation}\label{equ1}
	\begin{aligned}
		T_\text{avg}  & = \frac{T_\text{max}+T_\text{min}}{2}, \\
		T_\text{diff} & = T_\text{max}-T_\text{min}.
	\end{aligned}
\end{equation}
To proceed, we rely on the following equation:
\begin{equation}\label{equ1}
	\begin{aligned}
		\int^{T_\text{max}}_{T_\text{min}} e^{j2\pi ft} dt
		= & T_\text{diff}\text{sinc}\left(f T_\text{diff}\right)e^{j2\pi f T_\text{avg}},
	\end{aligned}
\end{equation}
where $\text{sinc}(x) = \frac{\sin(\pi x)}{\pi x}$. Then it is straightforward to reformulate (\ref{equ3}) as
\begin{align}\label{equ1}
    & \int^{\infty}_{-\infty} \phi_{l_1}(t)\phi^*_{l_2}(t-\tau) e^{-j2\pi\nu t} dt
    = T_\text{diff} 
    \\ & \cdot \text{sinc}\left\{ \left[(l_1-l_2)\Delta f -\nu\right] T_\text{diff} \right\}  e^{j2\pi \left\{\left[(l_1-l_2)\Delta f -\nu\right] T_\text{avg} + l_2\Delta f \tau \right\}  }. \nonumber
\end{align}
Accordingly, the AF may be expressed in compact form as 
\begin{equation}\label{equ1}
	\begin{aligned}
		\mathcal{X}(\tau,\nu) = \mathcal{X}_\text{Self}(\tau,\nu) +\mathcal{X}_\text{Cross}(\tau,\nu),
	\end{aligned}
\end{equation}
where
% \begin{equation*}
\begin{align}\label{equ1}
    & \  \mathcal{X}_\text{Self}(\tau,\nu) = T_\text{diff} \sum^{L-1}_{l=0} A^2_{l}  \text{sinc}\left( -\nu T_\text{diff} \right) \cdot e^{j2\pi (-\nu T_\text{avg}+ l\Delta f \tau)} \nonumber
    \\ & \mathcal{X}_\text{Cross}(\tau,\nu)  = T_\text{diff} \sum^{L-1}_{l_1=0} \sum^{L-1}_{l_2=0 \atop l_2\neq l_1}  A_{l_1}A_{l_2} e^{j(\psi_{l_1}-\psi_{l_2})} 
    \\ & \ \cdot \text{sinc}\left\{ \left[(l_1-l_2)\Delta f -\nu\right] T_\text{diff} \right\} \cdot e^{j2\pi \left\{\left[(l_1-l_2)\Delta f -\nu\right] T_\text{avg} + l_2\Delta f \tau \right\}  }. \nonumber
\end{align}
% \end{equation*}
The AF is composed of $\mathcal{X}_\text{Self}(\tau,\nu)$ and $\mathcal{X}_\text{Cross}(\tau,\nu)$, which are the superposition of $L$ self-AF components and $L(L-1)$ cross-AF components, respectively.

% On account of the randomness of OFDM communication signal, we analyze the statistical characteristics of its AF in the following.

\subsection{Statistical Characteristics of $\mathcal{X}_\text{Self}(\tau,\nu)$}
The expectation of $\mathcal{X}_\text{Self}(\tau,\nu)$ is irrelevant to the constellation probabilities, since $E\{A^2_l\}=1$ holds for any constellations. As a consequence, we mainly concentrate on the variance of $\mathcal{X}_\text{Self}(\tau,\nu)$, which can be derived as
% \begin{equation}
\begin{align}\label{equ8}
    & \sigma^2_\text{Self}(\tau,\nu) = E[|\mathcal{X}_\text{Self}(\tau,\nu)|^2] - |E[\mathcal{X}_\text{Self}(\tau,\nu)]|^2  \nonumber
    \\ & = T^2_\text{diff} \sum^{L-1}_{l_1=0}\sum^{L-1}_{l_2=0} E\left\{A^2_{l_1}A^2_{l_2}\right\}  \text{sinc}^2\left( - \nu T_\text{diff} \right) \cdot e^{j2\pi (l_1-l_2)\Delta f \tau}  \nonumber
    \\ & \quad - T^2_\text{diff} \sum^{L-1}_{l_1=0}\sum^{L-1}_{l_2=0}  \text{sinc}^2\left( - \nu T_\text{diff} \right) \cdot e^{j2\pi (l_1-l_2)\Delta f \tau}. 
\end{align}
% \end{equation}
For PSK, $E\left\{A^2_{l_1}A^2_{l_2}\right\}=1$ leads to $\sigma^2_\text{Self}=0$. In contrast, the variance for QAM is not zero, and is given as
\begin{equation}\label{equ9}
	E\left\{A^2_{l_1} A^2_{l_2}\right\} =\left\{
	\begin{aligned}
		& { E\left\{A^4_{l_1}\right\} }, \ & l_1=l_2 \\
		& E\left\{A^2_{l_1}\right\} \cdot E\left\{A^2_{l_2}\right\}=1, \ & l_1\neq l_2 
	\end{aligned}
	\right.
\end{equation}
Thanks to (\ref{equ9}), we can further simplify (\ref{equ8}) as
% \begin{equation}
\begin{align}\label{equ10}
    & \sigma^2_\text{Self}(\tau,\nu) = T^2_\text{diff} \text{sinc}^2\left( - \nu T_\text{diff} \right) \bigg\{ \sum^{L-1}_{l_1=0} E\left\{A^4_{l_1}\right\} \nonumber
    \\ &  + \sum^{L-1}_{l_1=0}\sum^{L-1}_{l_2=0,\atop l_2\neq l_1}   e^{j2\pi (l_1-l_2)\Delta f \tau}  
    - \sum^{L-1}_{l_1=0}\sum^{L-1}_{l_2=0} e^{j2\pi (l_1-l_2)\Delta f \tau} \bigg\} \nonumber 
    \\ & = T^2_\text{diff} \text{sinc}^2\left( - \nu T_\text{diff} \right)  \sum^{L-1}_{l_1=0} \left({E\left\{A^4_{l_1}\right\} - 1} \right).  
\end{align}
% \end{equation}
The above result suggests that the variance of the AF is mainly determined by the fourth moment of the constellation points, namely, $E\{A_l^4\}$. Moreover, it also clearly indicated that $\sigma^2_\text{Self}(\tau,\nu\neq 0) \ll \sigma^2_\text{Self}(\tau,\nu = 0) $. Therefore, the major impact of randomness lies in the zero Doppler slice $\sigma^2_\text{Self}(\tau,0)$, namely, the variance of autocorrelation function, in the form of
\begin{equation}\label{equ1}
	\begin{aligned}
        \sigma^2_\text{Self}(\tau,0)  = LT^2_\text{diff} \left(E\left\{A^4_{l}\right\} - 1 \right).
	\end{aligned}
\end{equation}
Note that the variance is always non-negative by definition. Hence, we have the following proposition.

$\mathrm{Proposition \ 1:}$ $E\left\{A^4_{l}\right\} - 1 \geq 0$.

$\mathrm{Proof:}$
Denote the number of constellation points by $Q$ and the $q$th probability by $p_q$. Owing to
$\sum^{Q}_{q=1}p_qA^2_q=1$
and
$\sum^{Q}_{q=1}p_q=1$,
we have
\begin{equation}\label{equ1}
	\begin{aligned}
		E\left\{A^4_{l}\right\} = & E\left\{A^4_{l}\right\} \sum^{Q}_{q=1}p_q =   \sum^{Q}_{q=1}p_qA^4_q \sum^{Q}_{q=1}p_q 
		\\ \geq & \left( \sum^{Q}_{q=1}\sqrt{p_q} A^2_q \sqrt{p_q} \right)^2 
		= \left( \sum^{Q}_{q=1}p_qA^2_q \right)^2 
		= 1.
	\end{aligned}
\end{equation}
The equal sign holds when $\frac{\sqrt{p_1} A^2_1}{\sqrt{p_1}}=\cdots=\frac{\sqrt{p_Q} A^2_Q}{\sqrt{p_Q}}$, i.e. $A^2_1=A^2_2=\cdots=A^2_Q {= 1}$, leading to unit modulus of all constellation points, i.e., PSK modulations.

% $\mathrm{Remark \ 2:}$ It is now evident that PSK-OFDM outperforms QAM-OFDM thanks to the zero variance of the former, i.e. 
% $$
% \sigma^2_\text{Self,QAM}(\tau,0) > \sigma^2_\text{Self,PSK}(\tau,0) = 0.
% $$

\subsection{Statistical Characteristics of $\mathcal{X}_\text{Cross}(\tau,\nu)$}
Similarly, the variance of $\mathcal{X}_\text{Cross}(\tau,\nu)$ can be expressed as
\begin{equation}\label{equ1}
	\begin{aligned}
		\sigma^2_\text{Cross}(\tau,\nu)=E\{\left|\mathcal{X}_\text{Cross}(\tau,\nu)\right|^2\} - \left|E\{\mathcal{X}_\text{Cross}(\tau,\nu)\}\right|^2.
	\end{aligned}
\end{equation}
By noting that 
\begin{equation}\label{equ1}
	\begin{aligned}
        E\{A_{l_1}A_{l_2} e^{j(\psi_{l_1}-\psi_{l_2})} \} = & E\{A_{l_1} e^{j\psi_{l_1}} \}E\{A_{l_2} e^{-j\psi_{l_2}} \} \\ = & 0, \ \ \ l_1\neq l_2,
	\end{aligned}
\end{equation}
$|E\{\mathcal{X}_\text{Cross}(\tau,\nu)\}|^2=0$ is obtained. This can be proved according to the symmetry of constellation points. 
Therefore, we only need to compute $E\{|\mathcal{X}_\text{Cross}|^2\}$, which is expressed as
% \begin{equation}
\begin{align}\label{equ13}
    & E\{|\mathcal{X}_\text{Cross}(\tau,\nu)|^2\} = T^2_\text{diff}  \sum^{L-1}_{l_1=0} \sum^{L-1}_{l_2=0,\atop l_2\neq l_1} \sum^{L-1}_{l'_1=0} \sum^{L-1}_{l'_2=0,\atop l'_2\neq l'_1} E\big\{ A_{l_1}A_{l_2}A_{l'_1}A_{l'_2} \nonumber
    \\ & \cdot e^{j(\psi_{l_1}-\psi_{l_2}-\psi_{l'_1}+\psi_{l'_2})}  \big\} \text{sinc}\left\{ 2\pi\left[(l_1-l_2)\Delta f -\nu\right] T_\text{diff} \right\}  \nonumber
    \\ & \cdot e^{j2\pi \left\{\left[(l_1-l_2)\Delta f -\nu\right] T_\text{avg} + l_2\Delta f \tau \right\} } \text{sinc}\left\{ 2\pi\left[(l'_1-l'_2)\Delta f -\nu\right] T_\text{diff} \right\} \nonumber
    \\ & \cdot e^{-j2\pi \left\{\left[(l'_1-l'_2)\Delta f -\nu\right] T_\text{avg} + l'_2\Delta f \tau \right\} }.
\end{align}
% \end{equation}
For further simplifications, it is evident that we only need to derive
\begin{equation}\label{equ160}
	\begin{aligned}	
	& E\left\{A_{l_1}A_{l_2}A_{l'_1}A_{l'_2} e^{j(\psi_{l_1}-\psi_{l_2}-\psi_{l_1}+\psi_{l_2})} \right\} 
	\\ & = \left\{
		\begin{aligned}
			E\left\{ A^2_{l_1}A^2_{l'_1} \right\}, \ \ l_1=l_2,l'_1=l'_2 \\
			E\left\{ A^2_{l_1}A^2_{l_2} \right\}, \ \ l_1=l'_1,l_2=l'_2 \\
			0, \quad \quad \quad \quad \quad \text{otherwise}
		\end{aligned}
		\right.
	\end{aligned}
\end{equation}

Note however that $\mathcal{X}_\text{Cross}(\tau,\nu)$ is defined when $l_2\neq l_1$ and $l'_2\neq l'_1$ in (\ref{equ13}). As a consequence, recalling (\ref{equ160}) demonstrates that all the non-zero components of $E\{|\mathcal{X}_\text{Cross}(\tau,\nu)|^2\}$ are contributed by the constraints of $l_1=l'_1$ and $l_2=l'_2$, yielding
\begin{equation}\label{equ1}
	\begin{aligned}
		\sigma^2_\text{Cross}(\tau,\nu) = & E\{|\mathcal{X}_\text{Cross}|^2\}-0  \\ = & T^2_\text{diff} \sum^{L-1}_{l_1=0} \sum^{L-1}_{l_2=0,\atop {l_2\neq l_1}}  E\left\{A^2_{l_1}A^2_{l_2} \right\}  \\ & \cdot \text{sinc}^2 \left\{ 2\pi\left[(l_1-l_2)\Delta f -\nu\right] T_\text{diff} \right\}. 
	\end{aligned}
\end{equation}
In addition, the condition of $l_2\neq l_1$ results in the independent random variables $A^2_{l_1}$ and $A^2_{l_2}$. Then we have
\begin{equation}\label{equ1}
	\begin{aligned}
		E \left\{A^2_{l_1}A^2_{l_2} \right\}  = E\left\{A^2_{l_1} \right\} \cdot E\left\{A^2_{l_2} \right\} = 1, \ {l_1\neq l_2}
	\end{aligned}
\end{equation}
Finally, the variance of $\sigma^2_\text{Cross}(\tau,\nu)$ is expressed as
\begin{equation}\label{equ1}
	\begin{aligned}
    \sigma^2_\text{Cross}(\tau,\nu) = & T^2_\text{diff} \sum^{L-1}_{l_1=0} \sum^{L-1}_{l_2=0,\atop {l_2\neq l_1}}  \text{sinc}^2 \left\{ 2\pi 
    \left[(l_1-l_2)\Delta f -\nu\right] T_\text{diff} \right\}.
	\end{aligned}
\end{equation}
When $\nu = 0$, $\sigma^2_\text{Cross}(\tau,\nu)$ can be approximately omitted owing to $\text{sinc}(2\pi(l_1-l_2)\Delta f)\approx 0$ for $l_1\neq l_2$. In contrast, $\sigma^2_\text{Cross}(\tau,\nu)$ may be relatively large when $\nu \neq 0$. However, $\sigma^2_\text{Cross}(\tau,\nu)$ is a deterministic value for each $\tau$ and $\nu$, which indicates that it is always constant for different constellations. 
As a consequence, we may only control $\sigma^2_\text{Self}(\tau,\nu)$ through constellation shaping.

%上式结论说明了以下问题：
%\begin{itemize}
%	\item $\sigma^2_\text{Cross}(\tau,0)$的值非常小，因为$\nu=0$并且$l_2\neq l_1$时，sinc函数的取值很小，也就是说$\mathcal{X}_{Cross}$对于自相关函数的影响很小；
%	\item 对于PSK或者QAM来说，$\sigma^2_\text{Cross}(\tau,\nu)$结果相同，也就是说\textbf{数据的随机性对$\mathcal{X}_\text{Cross}$的方差没有影响。}
%\end{itemize}

\section{PCS Method for ISAC}\label{sec3}
We now proceed to present the PCS-enabled signaling design method that allocates the probabilities of constellation points for OFDM ISAC signals, in order to control 
the statistical characteristics of AF. 
From the above analysis, and by recalling (\ref{equ10}), we naturally hope to devise a constellation that makes the fourth moment of random amplitude satisfy $E\{A^4_l\}=\sum^{Q}_{q=1}p_qA^4_q=c_0$, where $c_0$ is a preset parameter that controls the variance of the AF, in accordance with the system's requirements for S\&C. In light of Proposition 1, one should set $c_0\geq 1$. By doing so, $\sigma^2_\text{Self}(\tau,0)$ may be adjusted to control the performance tradeoff between S\&C. To realize this goal, we formulated the following optimization problem
\begin{equation}\label{equpcs}
	\text{(PCS)}\left\{
	\begin{aligned}
		\min_{\mathbf{p}} & \  \left|\sum^{Q}_{q=1}p_qA^4_q - c_0 \right|
		\\
		s.t. \ & \sum^{Q}_{q=1}p_qA^2_q = 1, \ \sum^{Q}_{q=1}p_q = 1 \\
		& 0 < p_q \leq 1,
	\end{aligned}
	\right.
\end{equation}
where $\mathbf{p} = [p_1, p_2,..., p_Q]^T$ represents the probability distribution vector.
In (\ref{equpcs}) we minimize the gap between the fourth moment of the constellation and a preset value $c_0$, subject to an average power constraint. 
It is readily to see that (\ref{equpcs}) is a convex optimization problem, which can be directly solved by CVX toolbox \cite{grant2014cvx}.
We also highlight that this operation is totally offline, 
which demonstrates that there is a consistent one-to-one match between each probability distribution of constellation points and each $c_0$.
As a consequence, such a method can be readily applied in practical base stations and user ends.

To evaluate the impact of $c_0$ on communications, we rely on the achievable information rate (AIR) in an AWGN channel $y=x+n$, where $y$, $x$, and $n$ denote the receive signal, the transmit data symbols modulated by arbitrary constellations, and the zero-mean Gaussian noise, respectively. The noise variance is denoted as $\sigma^2$. Note that this communication signal model can be viewed as a single channel case. For the transmitted OFDM signal with $L$ sub-carriers, the overall AIR is the superposition of AIRs from all sub-channels. 
It is known that in point-to-point (P2P) channels, the AIR is characterized by the input-output mutual information, which is expressed as
% \begin{equation}
\begin{align}\label{equ20}
    R_\text{sym} = & E_{X,Y}\left\{\log_2\frac{p_{Y|X}(y|x)}{{p_Y(y)}} \right\} \nonumber
    \\ = & \underbrace{\sum_{x} p(x) \int_{Y} \log_2 p(y|x) p(y|x) dy }_{-H(Y|X)=-\log_2(\pi e\sigma^2)} 
    \\ & \underbrace{- \int_{Y} \left[ \log_2 \sum_{x'} p(y|x') p(x') \right] \sum_{x} p(y|x)p(x)  dy }_{H(Y)}. \nonumber
\end{align}
% \end{equation}
Since $p_Y(y) = \sum_{x} p(y|x)p(x)$ is the sum of Gaussian probability density functions (PDFs) weighted by the prior probabilities of constellation points, the closed-form of $H(Y)$ cannot be obtained due to the Gaussian mixture PDF $p_Y(y)$. Instead, we approximately compute the entropy using Monte Carlo numerical integrals as follows:
\begin{equation}\label{equ45}
	\begin{aligned}
		H(Y) = & -E_Y \left[ \log_2 \sum_{x} p(y|x) p(x) \right] \\
		\approx & -\frac{1}{\text{MC}}\sum^{\text{MC}}_{k=1} \log_2 \sum_{x} p(y_k|x) p(x),
	\end{aligned}
\end{equation}
where $\text{MC}$ represents the number of Monte Carlo trials, $y_k$ denotes the $k$th observation and its conditional PDF $p(y_k|x)$ is with standard Gaussian forms in the $k$th trial, for each $x$ in the given constellation. By doing so, the entropy $H(Y)$ can be accurately approximated when $\text{MC}$ is sufficiently large.

\begin{figure}[!t]
	\centering
	\subfigure[Ambiguity function.] { \label{fig5a}
		\includegraphics[width=3.1in]{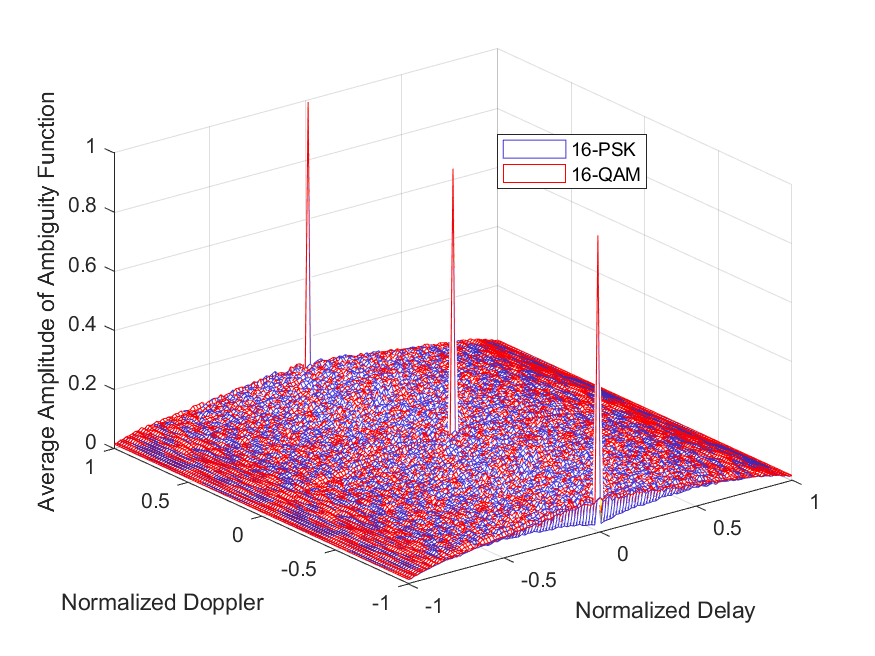}
	}	\\
	\subfigure[Zero Doppler slice.] { \label{fig5a}
		\includegraphics[width=1.6in]{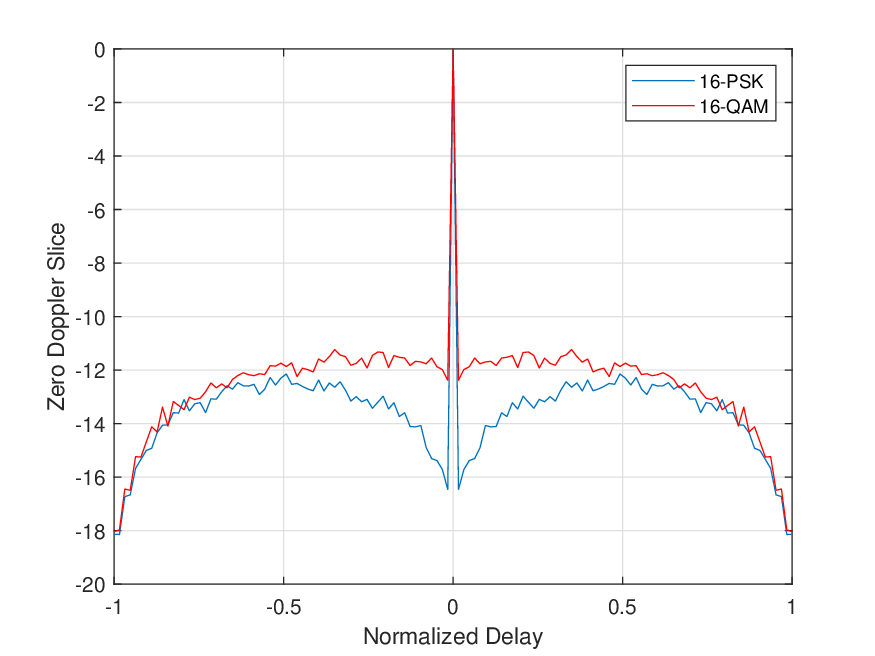}
	}
	\subfigure[Zero delay slice.] { \label{fig5b}
		\includegraphics[width=1.6in]{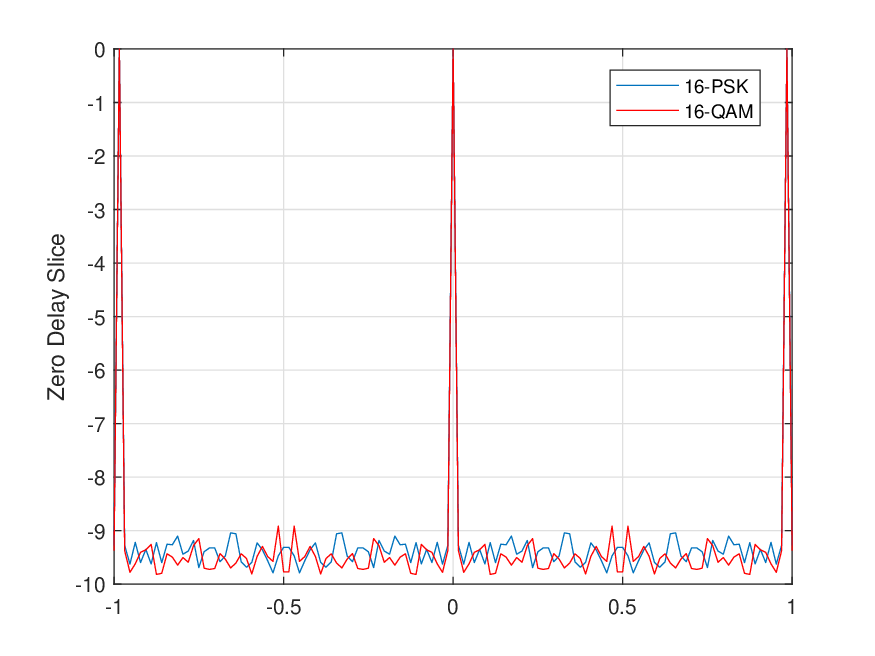}
	}
	\caption{Ambiguity function of 16-QAM and 16-PSK.}
	\label{fig1}
\end{figure}

\section{Simulations}
We consider an OFDM signal with $100$MHz bandwidth and $L=64$ subcarriers. Unless otherwise specified, we only adopt 16-QAM/PSK and 64-PSK/QAM for simulations, and designate the study of higher-order constellations as our future works. All the AFs are evaluated in accordance with their average AF performance, i.e. $\frac{1}{M}\sum^M_{m=1}|\mathcal{X}_m(\tau,\nu)|$, which experiences $M$ runs and $\mathcal{X}_m(\tau,\nu)$ represents the $m$th random realization of AF. In addition, the average AFs are normalized.

First, we evaluate the AF of the OFDM communication signal by illustrating the difference between 16-PSK and 16-QAM. As shown in Fig.\ref{fig1}, it is evident that 16-PSK exhibits significantly better performance compared to 16-QAM due to its lower sidelobes, with a maximum gap of 5dB, in the autocorrelation function. Additionally, the zero-delay slices of both 16-PSK and 16-QAM are nearly identical. It is worth noting that Fig.\ref{fig1}(a) displays three peaks, which is a result of using normalized axes and introduces two additional peaks caused by the Doppler ambiguity.

\begin{figure}[!t]
	\centering
	% Requires \usepackage{graphicx}
	\includegraphics[width=3.5in]{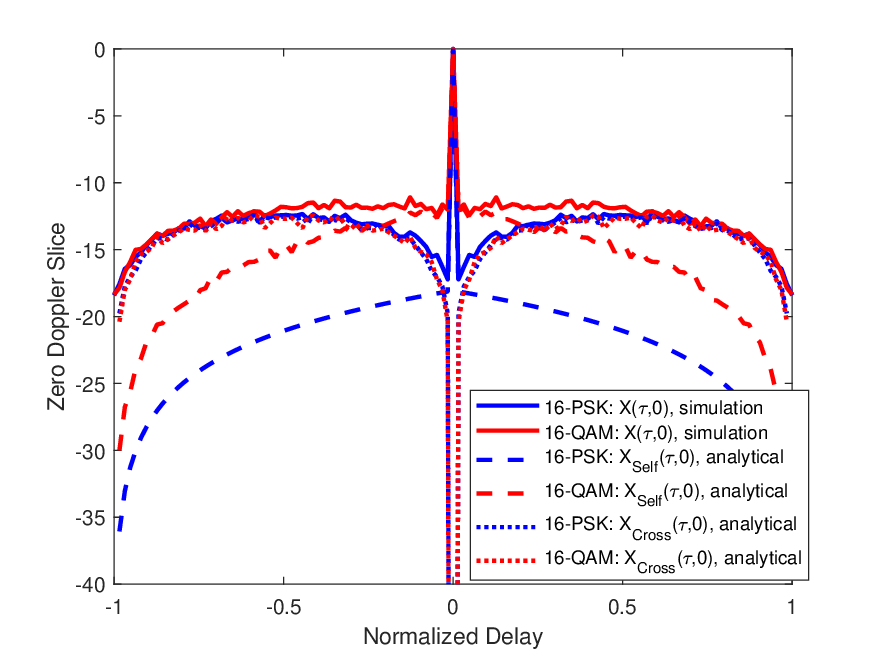}\\
	\caption{Autocorrelation function with simulated results $\mathcal{X}(\tau,0)$, and the analytical $\mathcal{X}_\text{Self}(\tau,0)$ and $\mathcal{X}_\text{Cross}(\tau,0)$.}
	\label{figx}
\end{figure}

Next, in Fig. \ref{figx}, we plot the analytical results of $\mathcal{X}_\text{Self}(\tau,0)$ and $\mathcal{X}_\text{Cross}(\tau,0)$, alongside the simulated autocorrelation functions of 16-QAM and 16-PSK. Evidently, the sensing performance gap between 16-QAM and 16-PSK stems solely from $\mathcal{X}_\text{Self}(\tau,0)$. Consequently, the statistical characteristics of $\mathcal{X}_\text{Cross}(\tau,0)$ bear no influence on the PCS method. This coincides with the analysis in Sec. \ref{sec2}.

\begin{figure}[!t]
	\centering
	% Requires \usepackage{graphicx}
	\includegraphics[width=3.5in]{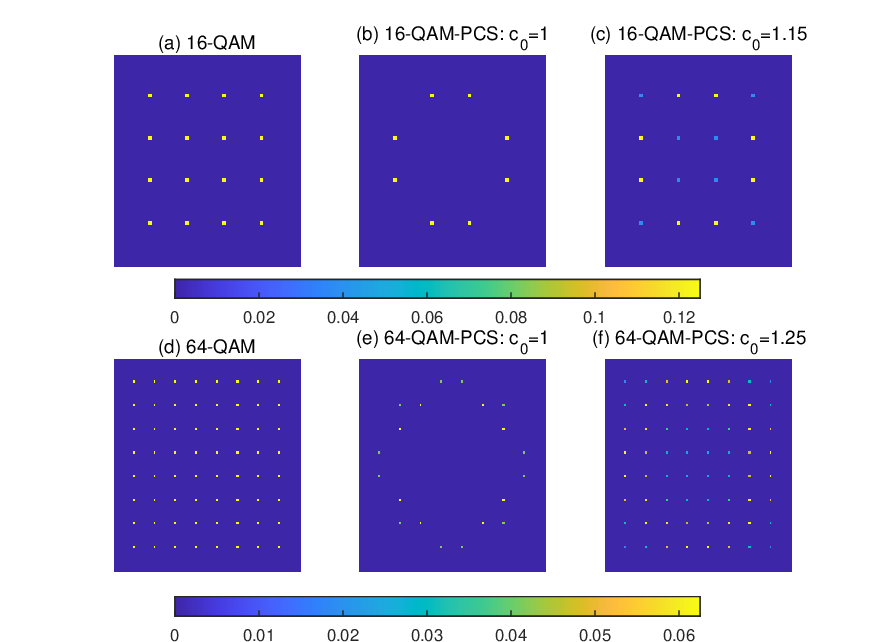}\\
	\caption{PCS results with different $c_0$: 16-QAM and 64-QAM. The brighter constellation point means its probability is larger.}
	\label{fig2}
\end{figure}

\begin{figure}[!t]
	\centering
	% Requires \usepackage{graphicx}
	\includegraphics[width=3.5in]{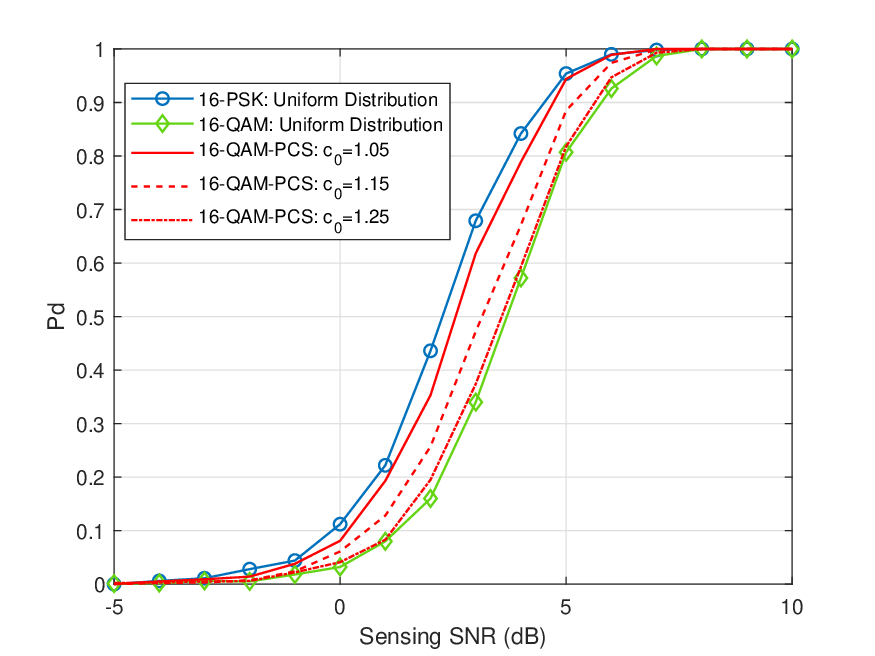}\\
	\caption{Probability of detection versus sensing SNR.}
	\label{fig3}
\end{figure}

Given a QAM-format constellation, we now test the performance of the proposed PCS approach (referred to as ``16-QAM-PCS''). As shown in Fig. \ref{fig2}, when $c_0=1$, the optimization model seeks the solution of best sensing performance. For 16-QAM, PCS outputs a unit modulus constellation, which is close to 8-PSK. Note that it is not a real 8-PSK since the angles between adjacent constellation points are not equal. For 64-QAM, PCS cannot find a constant modulus circle whose power is one, thereby outputs two constant modulus circles nearby the unit modulus circle. When $c_0$ increases, the PCS constellation results in deteriorated sensing performance. To illustrate this with a system-level simulation, we further consider a use case of detecting weak targets nearby the position of strong self-interference (SI), which is applicable for a practical scenario in full duplex radar sensing systems \cite{barneto2019full}. The smallest of constant false alarm probability (SO-CFAR) detector \cite{richards2014fundamentals} is exploited to address this problem, in order to exclude the SI from the computation process of detection threshold. 
Throughout 5000 Monte Carlo trials, the probability of false alarm is fixed as $10^{-4}$, and the weak target is within the $8$th range cell nearby the SI. 
The sensing signal-to-noise ratio (SNR) is defined as the power ratio between the weak target and the noise, while the power ratio between the SI and the noise is fixed as $10$ dB. Then the probability of detection (Pd) versus the sensing SNR is depicted in Fig. \ref{fig3}, which distinctly demonstrates that the practical sensing performance (i.e., the Pd) decreases with the increasing $c_0$.

\begin{figure}[!t]
	\centering
	% Requires \usepackage{graphicx}
	\includegraphics[width=3.5in]{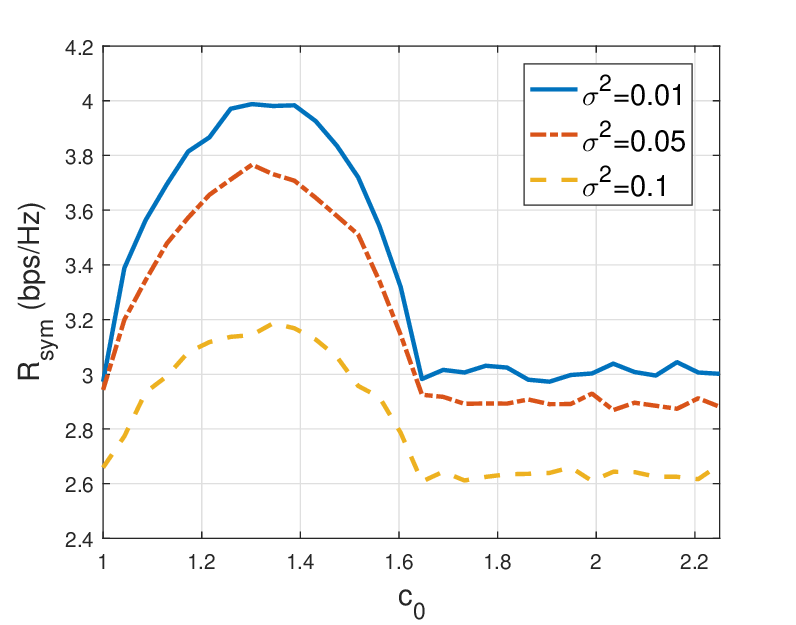}\\
	\caption{16-QAM-PCS: AIR versus $c_0$}
	\label{fig4}
\end{figure}

To evaluate the communication performance, we take 16-QAM as an example, and compute its AIR in an AWGN channel with Monte Carlo numerical integrals. In Fig. \ref{fig4}, $\sigma^2$ represents the power of noise, which controls the receive communication SNR. For a high SNR case ($\sigma^2=0.01$), it is revealed that AIR reaches the maximum value 4bps/Hz in $c_0=1.32$, which corresponds to the entropy of the uniformly distributed 16-QAM. When $c_0=1$, the AIR is 3bps/Hz in terms of the approximated 8-PSK constellation shown in Fig. \ref{fig2}(b), indicating that the best sensing performance is attained at the price of 1bit/Hz loss. 
Moreover, there is a distinct tradeoff between S\&C in the region of $c_0 \in [1,1.32]$, with known probability distributions of the constellation in this curve. Note that when $c_0>1.32$, the PCS is not uniform again, results in a declining AIR. When $c_0>1.62$, the fourth moment reaches to its largest value and thus AIR keeps constant as well.

\begin{figure}[!t]
	\centering
	% Requires \usepackage{graphicx}
	\includegraphics[width=3.5in]{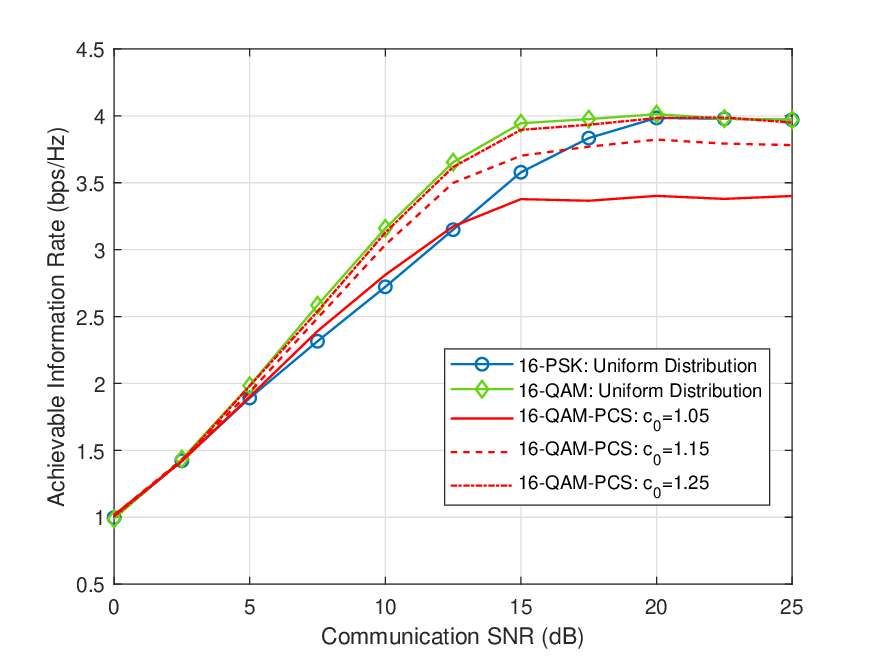}\\
	\caption{AIR versus communication SNR.}
	\label{fig6}
\end{figure}

Fig. \ref{fig6} further illustrates the advantages of the proposed PCS method for various SNR values. As anticipated, both 16-QAM and 16-PSK approaches reach their capacity limit of 4bps/Hz when the SNR is sufficiently high. However, in the relatively low SNR region, such as at SNR=10dB, a noticeable gap becomes apparent between 16-QAM and 16-PSK. Thanks to the PCS method, the optimized 16-QAM, i.e., 16-QAM-PCS, achieves a communication gain at the expense of the sensing performance loss compared to 16-PSK.

\section{Conclusion}
In this study, we proposed a novel PCS-enabled signaling design method to implement ISAC functionality using OFDM communication signals, while enhancing its sensing ability. By optimizing the derived fourth moment of constellation amplitudes, we are able to achieve a controllable and scalable tradeoff between sensing and communications, for OFDM signals modulated with random QAM symbols. Compared to the conventional uniformly distributed QAM, the proposed PCS-enbaled QAM attains better sensing performance. Meanwhile, compared to the conventional uniformly distributed PSK, our method improves the AIR in low communication SNR region.   
This offline operation demonstrates its potential for practical 6G ISAC applications.

Future work will be conducted from the following aspects:
\begin{itemize}
	\item The AF analysis may be generalized to the case of multiple OFDM symbols, rather than being restricted to a single symbol.
    \item The fundamental tradeoff of OFDM ISAC system in terms of pulse shaping and subcarrier power allocation may also be investigated.
	\item The achievable rate may be explicitly imposed in the PCS optimization problem as a constraint, which, however, needs to be solved via sophisticated numerical methods, e.g., the celebrated Blahut-Arimoto algorithm \cite{yeung2008information}.
\end{itemize}

%\section{Acknowledgement}
%The work was supported in part by the Natural Science Foundation of Jiangsu Province (No. BK20230416, BK20220438), the National Natural Science Foundation of China (No. 62301264, 62301268, 61971439, U22B2002), and the Startup Foundation for Introducing Talent of NUIST (No. 2023r018, 2023r015).

\ifCLASSOPTIONcaptionsoff
\newpage
\fi

\bibliographystyle{IEEEtran}
\bibliography{reference}

\end{document}